# In situ functionalization of graphene

*Kyrylo Greben[1,*], Sviatoslav Kovalchuk[1], Ana M. Valencia[2,4], Jan N. Kirchhof[1], Sebastian Heeg[1], Philipp Rietsch[3], Stephanie Reich[1], Caterina Cocchi[2,4], Siegfried Eigler[3] and Kirill I. Bolotin[1,*]*

[1] Department of Physics, Freie Universität Berlin, 14195 Berlin, Germany

[2] Department of Physics and IRIS Adlershof, Humboldt-Universität zu Berlin, 12489 Berlin, Germany

[3] Insititute of Chemistry and Biochemistry, Freie Universität Berlin, 14195 Berlin, Germany

[4] Carl von Ossietzky Universität Oldenburg, Physics Department, 26129 Oldenburg, Germany

*k.greben@fu-berlin.de

*kirill.bolotin@fu-berlin.de

**ABSTRACT**

While the basal plane of graphene is inert, defects in it are centers of chemical activity. An attractive application of such defects is towards controlled functionalization of graphene with foreign molecules. However, the interaction of the defects with reactive environment, such as ambient, decreases the efficiency of functionalization and makes it poorly controlled.

Here, we report a novel approach to generate, monitor with time resolution, and functionalize the defects *in situ* without ever exposing them to the ambient. The defects are generated by an energetic Argon plasma and their properties are monitored using *in situ* Raman spectroscopy. We find that these defects are functional, very reactive, and strongly change their density from $\approx 1\times10^{13}$ cm$^{-2}$ to $\approx 5\times10^{11}$ cm$^{-2}$ upon exposure to air. We perform the proof of principle *in situ* functionalization by generating defects using the Argon plasma and functionalizing them *in situ* using Ammonia functional. The functionalization induces the n-doping with a carrier density up to $5\times10^{12}$ cm$^{-2}$ in graphene and remains stable in ambient conditions.

**INTRODUCTION**

While the properties of pristine graphene are now largely understood, we are only beginning to understand the potential of controllably functionalized graphene. During the last decade, multiple approaches have been developed to attach foreign molecules such as hydrogen, oxygen, fluorine, or organic compounds to the basal plane of graphene[1–6]. Controlled functionalization has been used to open the band gap[7,8], adjust the doping levels[9], induce defect states producing photoluminescence[10–12], or perhaps even to induce magnetism in graphene[13].

Moreover, graphene controllably functionalized with biomolecules is in demand for applications in filtration, biotechnology, and biosensorics[14,15].

In general, there are covalent and non-covalent functionalization approaches[2,16,17]. In non-covalent functionalization, a target molecule is deposited onto graphene predominantly through interactions like van der Waals forces or π-π stacking[18]. As these interactions are relatively weak, molecules tend to cluster[19] or may be removed during processing of functionalized material[20]. In the covalent approach, a covalent bond forms between graphene and a target molecule. As the basal plane of graphene is highly inert, this functionalization approach requires reactive compounds, e.g. free radicals[2,3,20,21]. At the same time, defects in graphene are the centers of chemical activity. Therefore, many functionalization strategies use these defects to graft desired functionalities[22–28].

One of the most simple, cheap, and scalable techniques to induce defects in graphene is the exposure to an energetic plasma discharge[29]. The density, type, and configuration of defects can then be tuned by controlling the plasma type, energy, and exposure duration. However, in the majority of functionalization approaches, graphene is exposed to ambient before coming into contact with the target molecule[23,30–32]. As a result, freshly-created defects react with moisture, oxygen or hydrocarbons in the ambient reducing the efficiency and decreasing the control of functionalization[33,34]. This hinders the potential of plasma-treated graphene as the platform for controllably functionalized graphene-based hybrid materials.

Here, we overcome this problem by functionalizing freshly prepared plasma-induced defects in graphene without ever exposing them to the ambient. To accomplish this, we first explore the properties of plasma-induced defects in graphene. We show that these defects are functional rather than structural and that they are stable in vacuum but strongly react with the ambient. We then demonstrate a proof-of-principle functionalization of Ar plasma-induced seed-point defects with the $NH_3$ functional. We confirm functionalization by examining the evolution of carrier density, defect density, and strain extracted from time-resolved *in situ* Raman spectroscopy measurements.

## RESULTS

Our overarching goal is to develop an approach to controllably functionalize the basal plane of graphene. Towards this goal, we monitor the formation, study the properties, and functionalize defects in graphene *without exposing these defects to ambient*. To accomplish this, we have developed a setup that allows *in situ* 1) generation, 2) live monitoring, 3) annealing, and 4) functionalization of defects. The setup is a vacuum chamber with optical and gas access (Fig. 1a). Defects are generated in pristine monolayer CVD graphene by exposure to Ar or $NH_3$ plasmas, generated by radio frequency (RF) discharge. To characterize defect properties, the sample is continuously monitored *in situ* with Raman spectroscopy (Methods). Finally, plasma-generated defects can be functionalized using vapor deposition technique avoiding the exposure of the sample to ambient.

We use Raman spectroscopy to extract the defect density, carrier density, and strain in graphene as a function of time. The intensity, full width at half maximum (FWHM) and spectral positions of graphene Raman modes G and 2D (≈1591 $cm^{-1}$ and ≈2685 $cm^{-1}$, Fig. 1b,c) are used to gauge the initial graphene quality[35] and to extract carrier density and strain[36,37]. Disorder, such as structural defects (e.g., missing carbon atom) or $sp^3$-defects (e.g., attached organic molecules), activates the D mode as well as D' and D+D' modes in graphene (≈1594 $cm^{-1}$, ≈1625 $cm^{-1}$, and ≈2930 $cm^{-1}$, respectively)[35,38,39]. We use the ratio between the intensities of D and G modes to extract the density of defects introduced during the plasma exposure[40–42].

Our first goal is to investigate generation, stability, and reactivity of defects introduced in graphene via exposure to Ar plasma. At the beginning of the experiment, the sample is loaded into the vacuum chamber that is first pumped down to high vacuum ($p \approx 10^{-5}$ mbar) and then filled with the Ar gas at partial pressure p = 5 mbar (time t = 0). The Raman spectra are continuously acquired every five seconds (Fig. 1b). At t = 135 s, we generate defects igniting plasma for 5 seconds at -2dBm power. The sample is kept in medium vacuum, until we repeat the plasma exposure at t = 470 s for another 10 seconds at -2dBm power. The sample is further kept in medium vacuum until t = 1170 s to examine the stability of defects. Finally, at t = 1175 s the chamber is filled with air up to ambient pressure and monitored for ≈ 500 s after that.

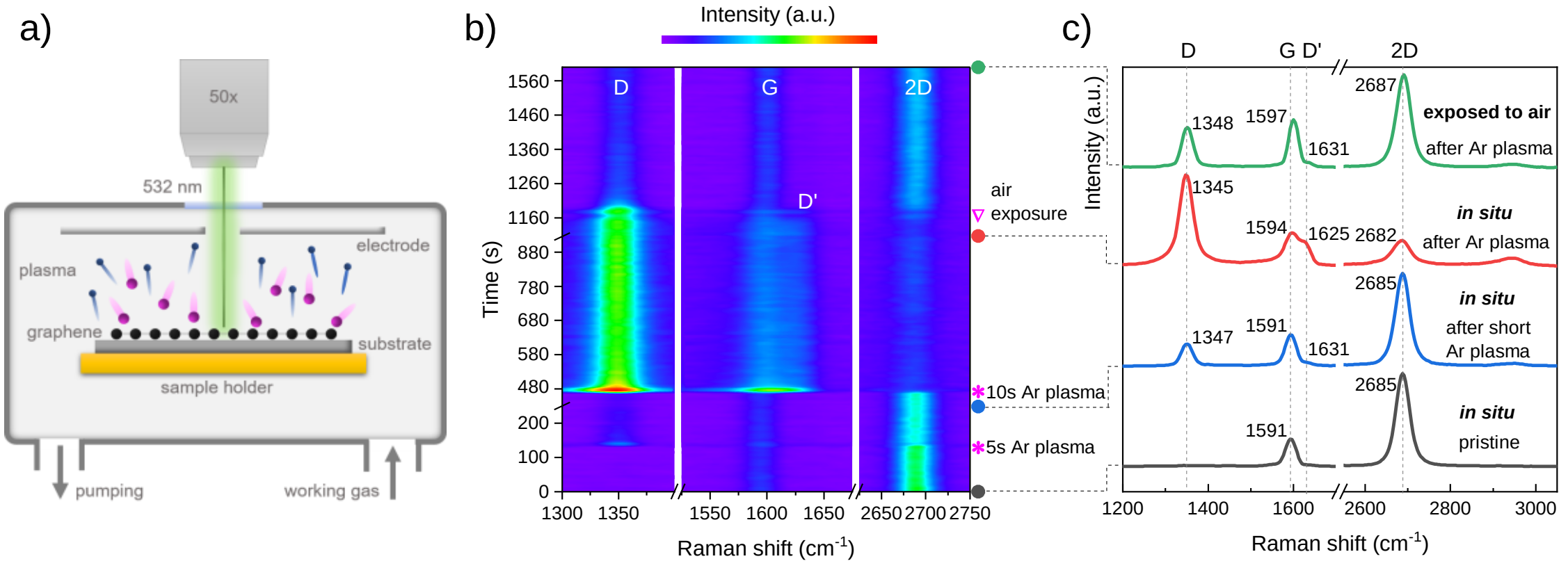

Figure 1: **Experimental setup and time-resolved Raman spectra.** a) Experimental setup for *in situ* generation/functionalization of defects and their monitoring via Raman spectroscopy. b) Time series showing the evolution of the graphene Raman spectrum. Until 135 s the sample is kept in vacuum. Argon plasma is ignited at 135 and 470 s, the sample is vented to air at 1175s. Time axis contains breaks. c) Several Raman spectra acquired at specific times marked in b).

We observe stark changes in the Raman spectrum during the entire process. At the beginning of the experiment, the ratio between 2D and G modes as well as the absence of the D mode indicate the negligible defect density in pristine CVD graphene (Fig. 1c, black). These spectra are uniform across the sample surface (Supplementary Fig. S1) and are stable over time. The first, five-second long plasma exposure introduces defects and activates the D mode in graphene (Fig. 1c, blue). The second plasma exposure changes the spectra dramatically: the intensity of a 2D mode strongly decreases, additional D' and D+D' modes appear, and the D mode further increases and begins to dominate the spectrum. All Raman modes shift and change relative intensities (Fig. 1c, red). Spectra remain relatively stable while the sample is kept in medium vacuum ($p_{Ar}$ = 5 mbar) between t = 520 s and 1170 s. As the sample is exposed to ambient at t ≈ 1175 s, the spectra change once again: the D mode decreases, the D' and D+D' modes almost completely disappear, and the 2D/G ratio goes back to its original value. Finally, after ≈ 100 s, the changes saturate and the spectra are relatively stable (Fig. 1c, green).

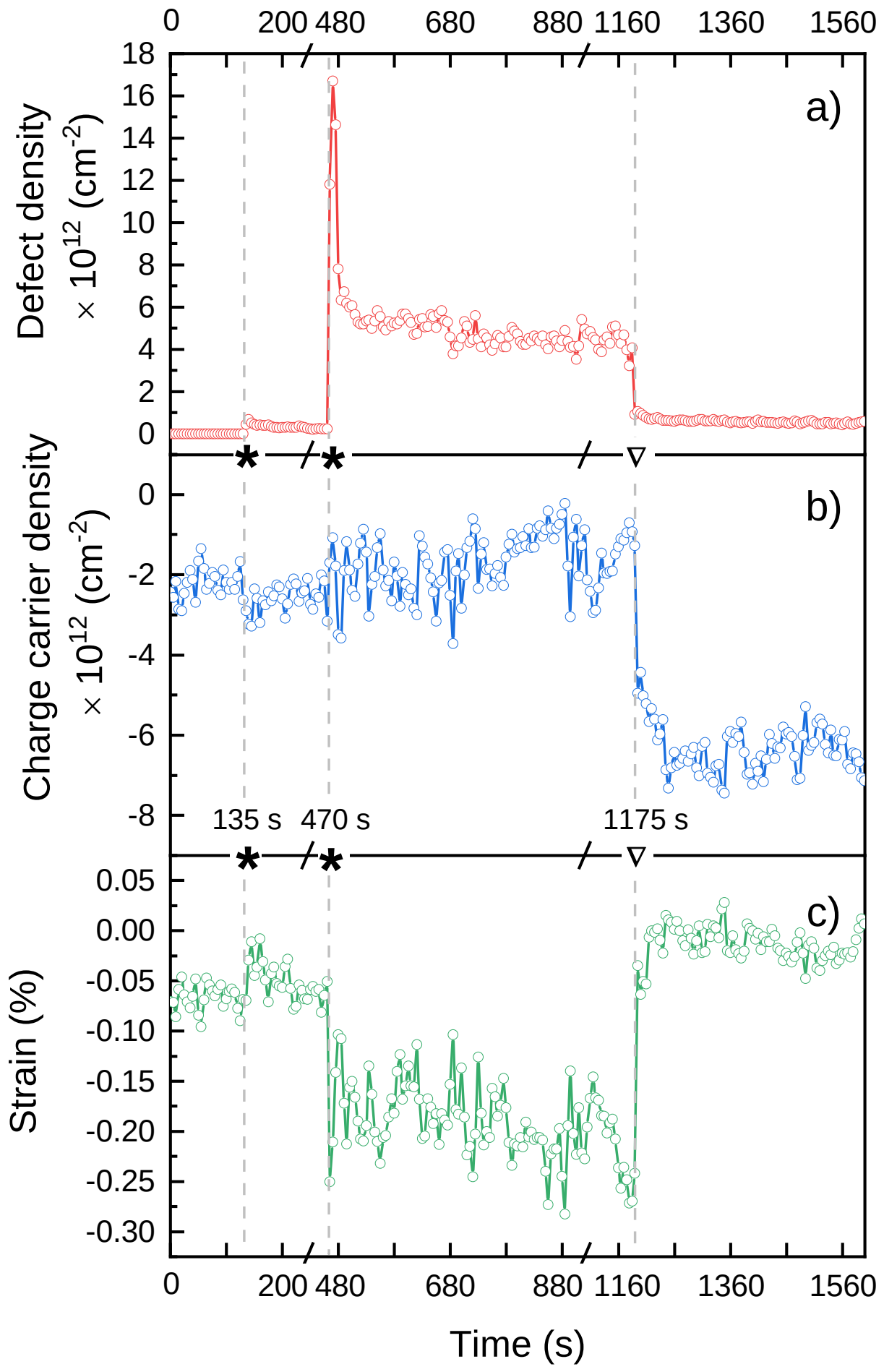

**Figure 2: Time-resolved changes in sample properties after plasma exposure and venting to air**. Time-dependent a) defect density, b) doping density and c) strain extracted from the Raman spectra with 5 sec. resolution using the procedure described in the text. The time axis is the same as in Fig. 1b.

To quantitatively examine modifications of graphene due to Ar plasma treatment and consecutive air exposure, in Figure 2 we extract the time-dependent defect density, charge carrier density, and strain of our sample during the entire experiment from the Raman data of Fig. 1b. We discuss the detailed analysis of the time-resolved Raman spectra in the Supplementary Information. We find that our graphene sample analyzed in Figs. 1 and 2 is initially p-doped and pre-strained (Supplementary Fig. S2).

At the beginning of the experiment the defect density is near-zero; pre-strain is low and initial carrier density is ≈ $-2.2\times10^{12}$ $cm^{-2}$, with minus sign corresponding to hole-doping. Both plasma exposure steps change defect density, doping, and strain. The defect density after the second discharge is ≈ $1.6\times10^{13}$ $cm^{-2}$ and rapidly (≈ 50 s) decreases to ≈ $6\times10^{12}$ $cm^{-2}$ after the plasma is turned off. Plasma exposures induce n-doping of ≈ $7\times10^{11}$ $cm^{-2}$ and strain of ≈ 0.1%.

After the fast dynamic following the plasma exposures, the sample is stable in Argon ($p_{Ar}$ = 5 mbar) as the carrier density, strain, and defect density remain stable in the interval t = 520 – 1170 s. At the time t = 1175 s, we start filling the chamber with air. We observe a rapid decrease of the defect density by an order of magnitude, to ≈ $5\times10^{11}$ $cm^{-2}$ within 40 s (Fig. 2a). Simultaneously, we observe p-doping from air exposure, ≈ $5\times10^{12}$ $cm^{-2}$ (Fig. 2b), accompanied by the relaxation of strain (Fig. 2c). We note that changes in the carrier density affect the intensity ratio I(D)/I(G) used to extract the defect density[43,44]. These effects are accounted for in the analysis of Fig. 2 (Supplementary Information). Following these initial fast changes, we observe slow dynamics on the time scale of hours. During that time, the defect density decreases by more than a factor of two and the carrier density increases by an order of magnitude (Supplementary Fig. S5).

To summarize our observations so far, the data of Figs. 1–2 show that *in situ* plasma-induced defects in graphene are stable in Argon but react with air. The density of these defects decreases by an order of magnitude from ≈ $6\times10^{12}$ $cm^{-2}$ in Argon to ≈ $5\times10^{11}$ $cm^{-2}$ in air. However, the question remains: what is the chemical/physical nature of these defects?

In general, the defects produced by plasma exposure[29] can be structural (i.e. missing carbon atom)[24] or functional ($sp^3$-like defects interacting with an external atom/molecule)[2,32]. These defect types are distinguished by their energy and related stability. To estimate this energy scale, we thermally anneal our samples. Figure 3 shows the evolution of the Raman spectra and calculated defect density for the sample annealed *in situ* in vacuum right after the introduction of defects. The D mode in Fig. 3a almost completely disappears after a relatively mild annealing at 85 °C, and the apparent defect density drops to the same value as in pristine graphene (Fig. 3b). This suggest that the defects produced by Ar plasma in our experiment *in situ* are functional and not structural. It is known that structural defects (missing carbon atoms) are stable up to much higher temperatures of 800–900 °C[45,46].

To figure out the type of functional attached to the carbon atoms, we performed DFT calculations of binding energies, induced doping, and strain for H-, OH- and O- functional defects (Supplementary Fig. S6). The lowest binding energy of -0.839 eV as well as induced electron doping and strain below 0.2% qualitatively suggest hydrogen as the most likely defect type induced by Ar plasma *in situ* at mbar pressures. Indeed, a similar behavior was observed for weakly bound functional defects in hydrogenated graphene[47–49]. In addition, hydrogen functionalities are expected to produce charge transfer and electron-doping[4,50] similar to the one observed in Fig. 2b as well as induce significant strain[51] due to modification of bond lengths, the behavior is seen in Fig. 2c. Finally, while the C-H bond is strong in bulk compounds, it is much more reactive in the case of graphene[4,33,52]. Therefore, it is not surprising that H-functionalities are removed from graphene upon exposure to ambient.

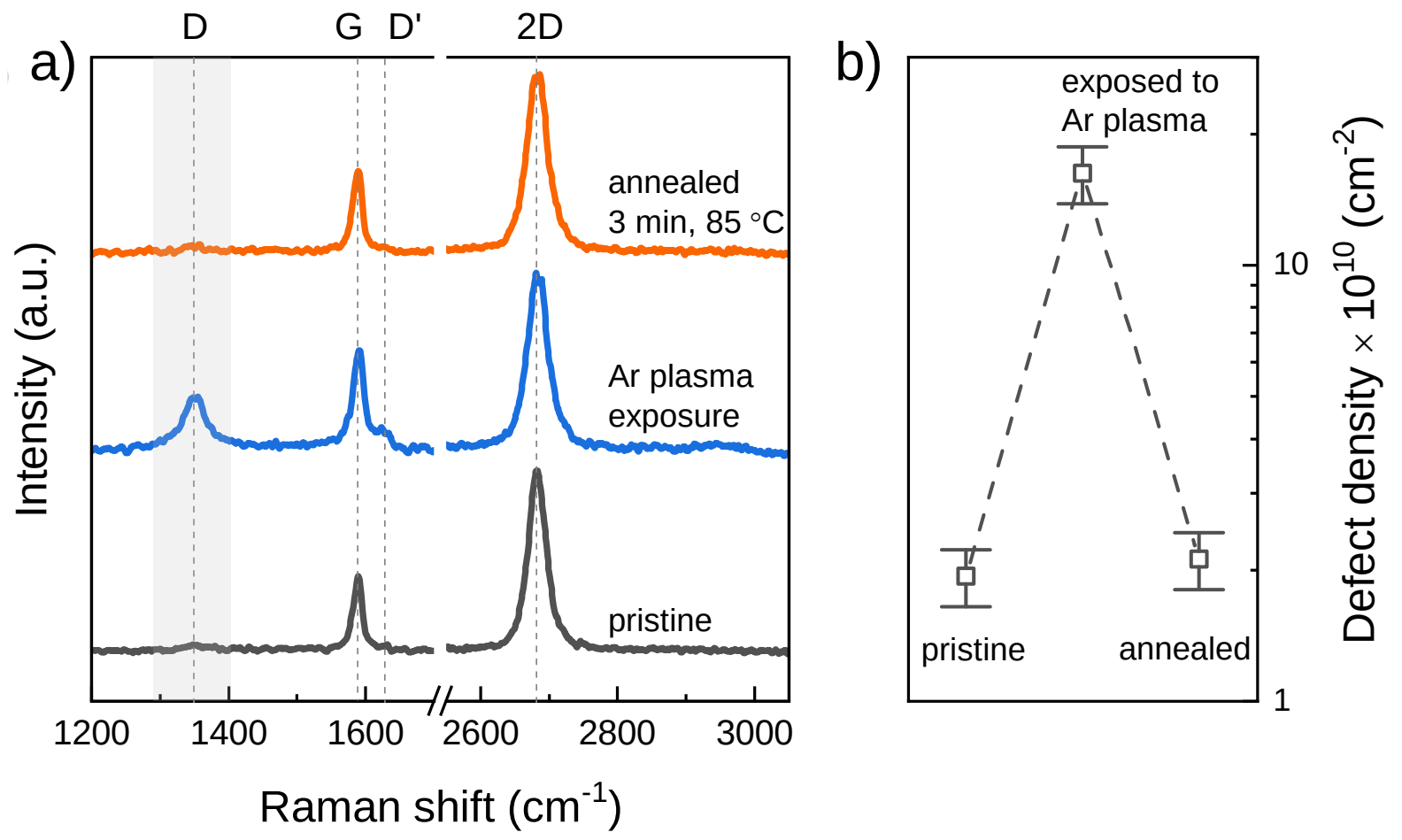

Figure 3: **Reversibility of plasma-induced defects upon *in situ* annealing. a)** Evolution of Raman spectra of graphene. A defect mode appears in pristine graphene (black) upon plasma exposure (blue). After *in situ* annealing to 85°C the mode disappears (red). **b)** The defect density extracted for each step in a).

There are two possible mechanisms for hydrogen functionalization in our experiments. First, $H_2$ that is present in trace concentrations in our chamber in medium vacuum becomes ionized together with Ar due to similar ionization energies[48] and may react with graphene[26,48]. Second, water adsorbed on our pristine samples may dissociate under ion/electron bombardment[53]. This could also lead to hydrogen functionalization. We note that more precise analytical techniques such as *in situ* XPS may distinguish between the proposed scenarios.

One particularly attractive application of reactive plasma-induced functional defects is for the further controlled chemical functionalization of graphene. Hydrogenated graphene is an interesting candidate for further chemical functionalization due to its reactivity[4,27,33,52]. The results above show that plasma-induced defects in graphene react with air. This greatly reduces their density and limits the *ex situ* functionalization potential. To overcome this limitation, we propose a new *in situ* functionalization pathway. The idea behind the approach is to introduce target molecular species into a vacuum chamber with freshly *in situ* Ar-induced functional defects *before the defects react with air*. We expect that the target species should attach to a large density of "seed-points" in graphene while these defects are still reactive. In the rest of the paper, we show the proof-of-principle of such two-step functionalization process.

To demonstrate the viability of our approach, we chose ammonia ($NH_3$) as our target functional. The interaction of ammonia with graphene is well understood and is commonly used to introduce a large carrier density in graphene[30,31,54], e.g. for applications in transparent conductive electrodes. In a proof-of-principle experiment, we first generated defects using Ar plasma as discussed above (10 s, 1 dBm, 0.1 mbar). In the second step, without breaking the vacuum, we introduced $NH_3$-plasma (15s, 1dBm, 0.2mbar) to functionalize the defects created during the first step. Finally, the sample was exposed to the ambient. Defect density and charge carrier density at each step of the functionalization process are shown in Fig. 4 (red points). For comparison, in the same graph we show a sample that was exposed to Ar plasma only (15 s, -2 dBm, 5 mbar, green points) and another sample that was exposed to $NH_3$ plasma only (50 s, 2 dBm, 0.2 mbar, blue points).

We first examine reference Ar-only and $NH_3$-only samples. In the Ar-only sample, as discussed above in Figs. 1 and 2, we created the defect density of $\approx 6\times10^{12}$ cm$^{-2}$, which induced a slight n-doping of $\approx 7\times10^{11}$ cm$^{-2}$ (Fig. 4, green points). This defect density drops by more than one order of magnitude upon exposure to ambient. In contrast, $NH_3$-plasma exposure generates, by itself, a large n-doping of $\approx 7\times10^{12}$ cm$^{-2}$, while generating the defect density of $\approx 2\times10^{11}$ cm$^{-2}$ (Fig. 4, blue points). After exposure to ambient, the concentration of defects is only slightly reduced, while the doping is reduced strongly. Similar results for $NH_3$ samples have been reported previously[29–31,54]. We conclude that both plasma treatments induce functional defects with different functional groups. The functional groups produced by Ar plasma (likely hydrogen functionalities) induce electron doping and appear to be reactive. In contrast, the groups produced by the $NH_3$ plasma (ammonia) induce electron doping and do not interact with ambient air strongly. The hole doping seen in both samples upon air exposure likely results from adsorption of water from ambient.

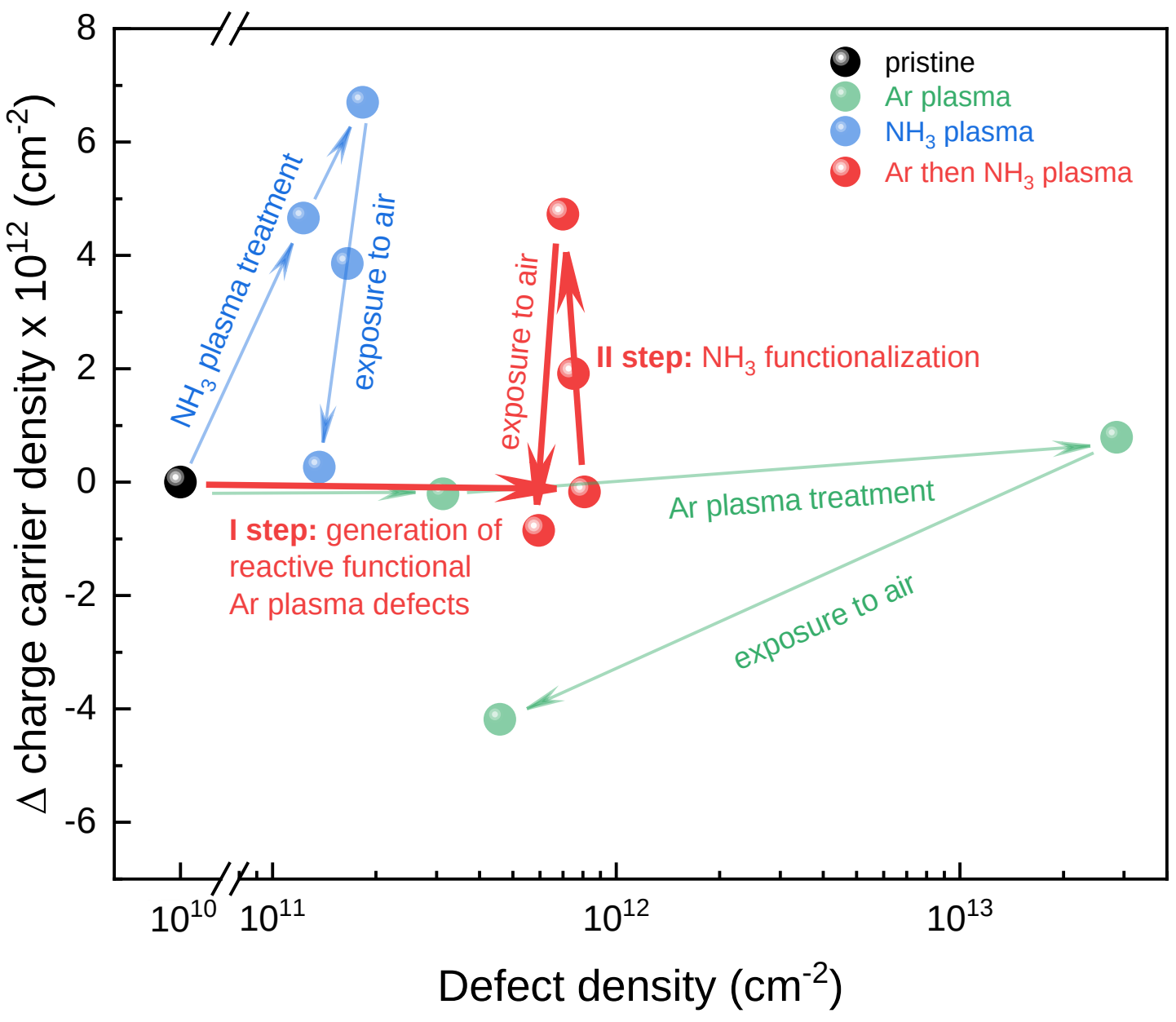

Figure 4: ***In situ* two-step functionalization of graphene.** The carrier density and the defect density shown for each step of the two-step functionalization approach of graphene (red). In that approach, anchor points are created via Argon plasma exposure and are functionalized by exposure to $NH_3$ plasma. For comparison, the samples exposed just to Ar plasma (green) and just to $NH_3$ plasma (blue) are also shown. The carrier density shown here is relative to the pristine state, to ease the comparison between the samples.

Finally, we examine the sample exposed to the two-step *in situ* functionalization process (Fig. 4, red). The first Ar plasma treatment results in the defect density $\approx 8\times10^{11}$ cm$^{-2}$. The following exposure to $NH_3$ plasma during the second step does not change the extracted defect density. Despite that, the carrier density increases to $\approx 5\times10^{12}$ cm$^{-2}$. Importantly, the defect density remains near constant upon exposure to ambient. All of that suggests that during the second functionalization step $NH_3$ derivatives bind to the reactive functional "seed-point" defects in graphene produced by Ar plasma in the first functionalization step rather than simply attach directly to graphene. Indeed, if latter was the case, we would expect to see an increase in the defect density upon $NH_3$ plasma exposure in the second step. In addition, the stability of the defect density in the two-step process suggests that functionalization of the defects is stable, unlike the case we observed for Ar plasma, but similar to what we have seen for $NH_3$ plasma. Finally, large electron doping after the two-step process suggests efficient $NH_3$ functionalization. All of that constitutes the proof of principle for our functionalization strategy.

Utilizing *in situ* functionalization method used here, other organic or inorganic functional can be introduced to graphene[27,33]. The advantage of this approach is the possibility to create high densities of "freshly-generated" reactive defects that could generate large doping of $> 10^{13}$ cm$^{-2}$, facilitate close packing of molecules, and allow the functionalization of graphene with previously inactive reagents.

In summary, we developed a new *in situ* approach to generate and monitor defects in graphene. We have shown that defects in graphene created via Ar plasma exposure are stable in vacuum but react with the ambient. Both the defect density and the carrier density in graphene decrease by about an order of magnitude upon exposure to ambient. We demonstrated a two-step *in situ* functionalization of graphene. In this process, we functionalized graphene with $NH_3$ functional at high density utilizing the reactive "seed-point" defects created via Ar plasma without exposure to ambient. We confirmed the functionalization by continuously analyzing defect density, carrier density, and strain in our samples through *in situ* Raman spectroscopy. Overall, we believe that our novel *in situ* functionalization approach using reactive defects opens the possibility to introduce various chemical functionalities to graphene and thereby providing a pathway towards scalable creation of various hybrid organic/inorganic 2D materials.

## METHODS

**Sample synthesis:** Single layer graphene is synthesized on the copper substrate by chemical vapor deposition (CVD). The mixture of methane (5 sccm), hydrogen (10 sccm), and argon (5 sccm) is let into the CVD chamber, which is kept at 1035 °C. The growth time is 7 min. After the growth, graphene is transferred onto the Si/$SiO_2$ substrate by a standard method[55].

**Setup:** The vacuum chamber is pumped down to $p \approx 10^{-5}$ mbar. The working gas (Ar or $NH_3$) is let into the chamber with the partial pressures of 0.1 – 5 mbar. The sample is located at the sample holder halfway between the electrode and the bottom of the chamber. The sample holder is electrically contacted for *in situ* annealing purposes. The plasma is generated via capacitive coupling of a plate electrode and the chamber using the microwave signal from HP8648B microwave generator at a constant frequency of 13.56MHz amplified by 50dB with the amplifier. The concentration of defects in graphene can then be controlled by adjusting the discharge power and plasma exposition time. The sample is monitored with *in situ* Raman spectroscopy in a modified Witec Alpha setup using 532 nm excitation wavelength.

**DFT calculations:** DFT calculations are carried out with the all-electron code FHI-aims[56]. Geometry optimization is performed within the generalized gradient approximation for the exchange-correlation functional using the Perdew-Burke-Ernzerhof parametrization[57]. Van der Waals interactions are included with the Tkatchenko-Scheffler scheme[58]. We employ tight integration grids and TIER2 basis sets[59], and the atomic positions are relaxed until the Hellmann-Feynman forces are smaller than $10^{-3}$ eV/Å.

## ACKNOWLEDGEMENTS

We gratefully acknowledge Dr. Georgy Gordeev for useful discussions. This work was supported by the Deutsche Forschungsgemeinschaft (DFG) - Projektnummer 182087777 - SFB 951 and ERC Starting grant no. 639739.

Supplementary Information for

# "In situ functionalization of graphene"

*Kyrylo Greben[1,*], Sviatoslav Kovalchuk[1], Ana M. Valencia[2], Jan N. Kirchhof[1], Sebastian Heeg[1], Philipp Rietsch[3], Stephanie Reich[1], Caterina Cocchi[2,4], Siegfried Eigler[3] and Kirill I. Bolotin[1,*]*

[1] Department of Physics, Freie Universität Berlin, 14195 Berlin, Germany

[2] Department of Physics and IRIS Adlershof, Humboldt-Universität zu Berlin, 12489 Berlin, Germany

[3] Insititute of Chemistry and Biochemistry, Freie Universität Berlin, 14195 Berlin, Germany

[4] Carl von Ossietzky Universität Oldenburg, Physics Department, 26129 Oldenburg, Germany

*k.greben@fu-berlin.de

*kirill.bolotin@fu-berlin.de

## SPATIAL HOMOGENEITY AND MEASUREMENT ERROR

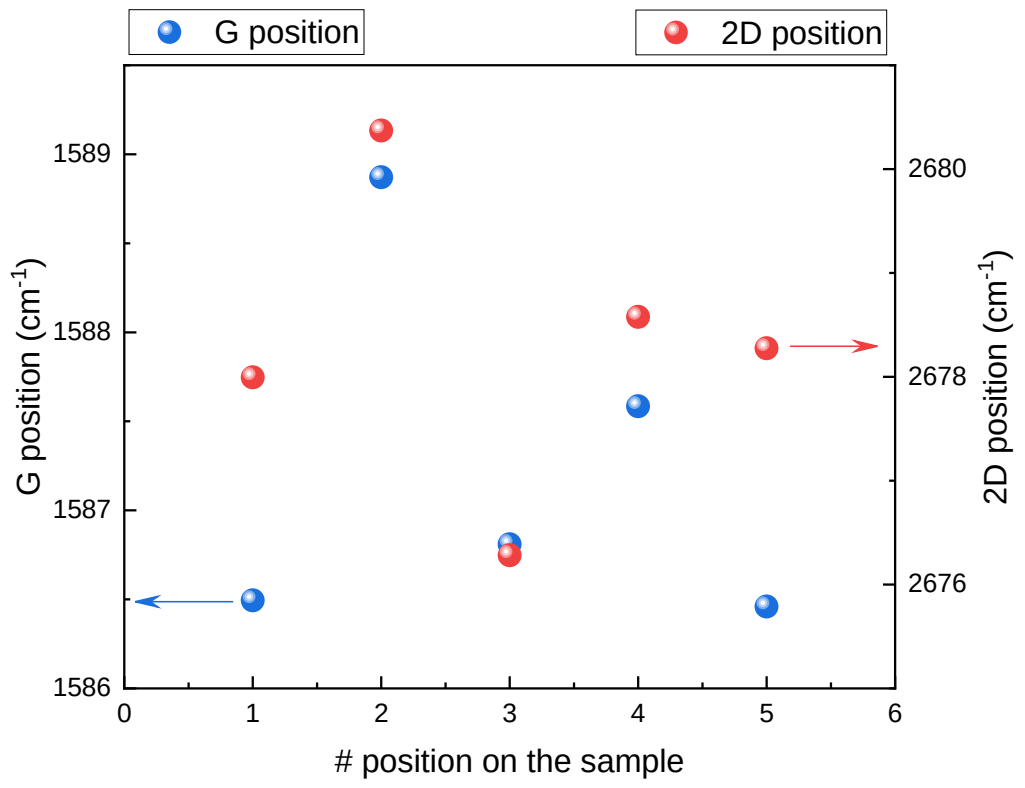

***Supplementary Figure S3: Spatial homogeneity of graphene and the estimation of a spatial error.*** *The spatial distribution of G (blue) and 2D (red) mode positions over the sample results in a standard deviation of 1.0 cm$^{-1}$ and 1.5 cm$^{-1}$ for the G mode and for the 2D mode positions, respectively. In the main text, we, therefore, accept 1.5 cm$^{-1}$ as the standard deviation when evaluating the position.*

## ANALYSIS OF RAMAN SPECTRA

We perform the analysis of the time-dependent Raman spectra from Fig. 1b of the main text in order to extract time-dependent defect density, charge carrier density, and strain for different stages of our experiment. We fit the Raman spectra obtained from the experiment to extract spectral frequencies, full width at half maximum (FWHM) and intensity for D, G, D′ and 2D modes. We calculate the defect density as follows. We first evaluate the ratio of intensities of the D and G modes, $I_D/I_G$. Then, using the laser wavelength ($\lambda$), we relate $I_D/I_G$ to the distance between defects ($L_D$), and then to the defect density $n_D$, using Eq. 1, as was introduced in detail in the Supplementary Reference S1:

$$n_D = \frac{10^{14}}{\pi L_D^2} = \frac{10^{14}}{\pi \times 1.3\times10^{-9}\lambda_L^4}\frac{I_D}{I_G} = \frac{4.14\times10^{22}}{532^4}\frac{I_D}{I_G}\ [cm^{-2}]. \qquad (1)$$

We note that for large defect densities (> $10^{12}$ $cm^{-2}$), Eq. 1 changes to

$$n_D = \frac{10^{14}}{\pi L_D^2} = \frac{4.14\times10^{22}}{532^4}\left(\frac{I_D}{I_G}\right)^{-1}\ [cm^{-2}]. \qquad (2)$$

In order to extract the charge carrier density and strain we follow the procedure outlined in Supplementary Refs. S2 and S3. We first plot the frequency of the 2D mode (Supplementary Fig. S2a) as well as G FWHM (Fig. S2b) vs. the frequency of the G mode for sample A, discussed in the main text. The values for the frequencies of the G and 2D modes, obtained from the experiment, are calibrated using the Ne lamp and the Ar plasma luminescence lines. The color scale represents the time from the beginning of the experiment (see main text). Dashed lines in Fig. S2a represent the influence of strain with the slope of 2.2 (black)[S2,3], p-doping with the slope of 0.55 (red)[S4], while the blue line represents the experimental data for n-doping from Ref. S4. The positions for unstrained and undoped graphene ($\omega_G$ = 1583 $cm^{-1}$ and $\omega_{2D}$ = 2678 $cm^{-1}$, black circle) are taken from Supplementary Refs. S2,4.

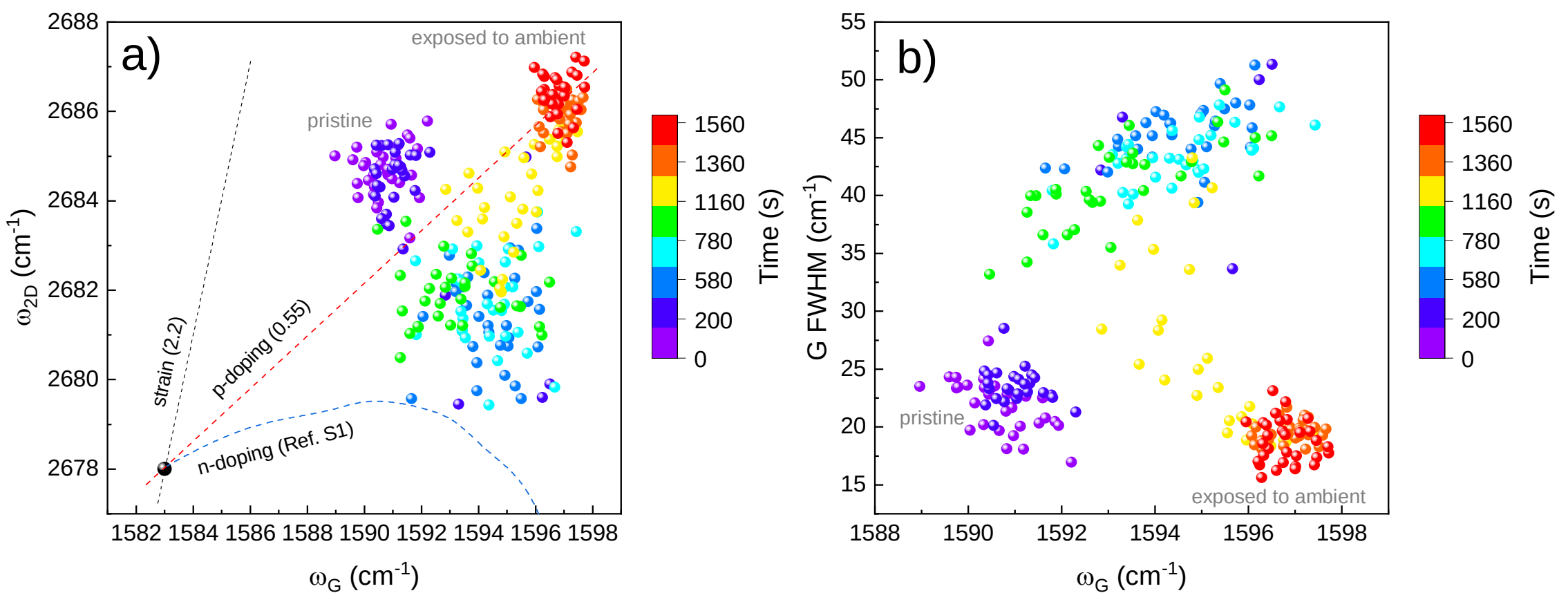

***Supplementary Figure S4: Time-resolved analysis of the Raman spectra from Fig. 1b in the main text. a)*** *$\omega_G$ vs. $\omega_{2D}$ mode plot and* ***b)*** *$\omega_G$ vs. G FWHM plot. The disordered state of the sample (after Ar plasma treatment) has a broader FWHM and is shifted due to the presence of disorder*[S1,5,6].

In order to obtain absolute values for the charge carrier density and the strain from the Supplementary Fig. S2, we use the black dashed line (strain) vs. red dashed line (p-doping) or blue dashed line (n-doping) as the axis of the coordinate system. We chose p-doping, because our pristine graphene is closest to the p-doping line and because graphene samples that have been exposed to ambient before the measurement are typically p-doped. In the strain-doping coordinate system, we find positions for each experimental point relative to the undoped and unstrained graphene. This is how the strain effect is separated from the doping effect[S2,3]. We then map the change in strain and doping each back to the shifts of the G-mode frequency ($\Delta\omega_G$). We use $\Delta\omega_G$ to obtain absolute values of the strain, under the assumption of the uniaxial strain[S3,7], using the Eq. 3:

$$\Delta\varepsilon = -\frac{\Delta\omega_G}{23.5}\ [\%]. \qquad (3)$$

Together with the Fermi velocity ($v_F$) as well as the analysis from the Supplementary Ref. S4 we obtain the absolute values for the charge carrier density, using the Eq. 4:

$$|n| = \pi\left(\frac{-18\Delta\omega_G - 83}{sgn(n)\hbar v_F}\right)^2\ [cm^{-2}]. \qquad (4)$$

We note that the Eq. 4 is valid for pre-doped samples[S4], where $E_F \lesssim -100\ meV$.

In order to obtain correct values for charge carrier density and strain it is important to account for all possible mechanisms that can influence the frequencies of the G- and 2D-modes. In addition to initial pre-strain and pre-doping, which we have already taken into account, disorder[S1,5,6] may additionally influence the frequencies of the G- and 2D-modes. Martins Ferreira et al. have experimentally measured[S5] the shift of the G, 2D and other modes' frequencies due to the disorder induced in graphene. This effect becomes relevant at high defect densities (> $10^{12}$ $cm^{-2}$). The disorder also strongly influences the G FWHM (Fig. S2b). When the doping is constant, the high defect densities (> $10^{12}$ $cm^{-2}$) lead to G FWHM > 35 $cm^{-1}$ and shifts of the G- and 2D-mode frequencies. The second Ar plasma exposure leads to the G FWHM > 35 $cm^{-1}$, shown in the Supplementary Fig. S2b. For these data points, the effect of disorder on the G- and 2D-mode frequencies has to be taken into account, and is performed as follows.

From the $\Delta\omega_G$ vs. $L_D$ and $\Delta\omega_{2D}$ vs. $L_D$ dependencies in the Supplementary Ref. S5 we extract the average $\Delta\omega_G$ and $\Delta\omega_{2D}$ of the two experimental points at each $L_D$. We then subtract the resulting disorder-induced shifts of the $\omega_G$ (e.g., 2.05 $cm^{-1}$ at $L_D \approx 2.2$ nm) and $\omega_{2D}$ (e.g., -8.59 $cm^{-1}$ at $L_D \approx 2.2$ nm) from the data in Fig. S2a. The resulting $\omega_{2D}$ vs. $\omega_G$ plot is shown in the Supplementary Fig. S3. Now, having corrected the absolute frequencies of the G- and 2D-modes for the effect of disorder, we can apply the analysis described above to extract the absolute charge carrier density and strain from our data.

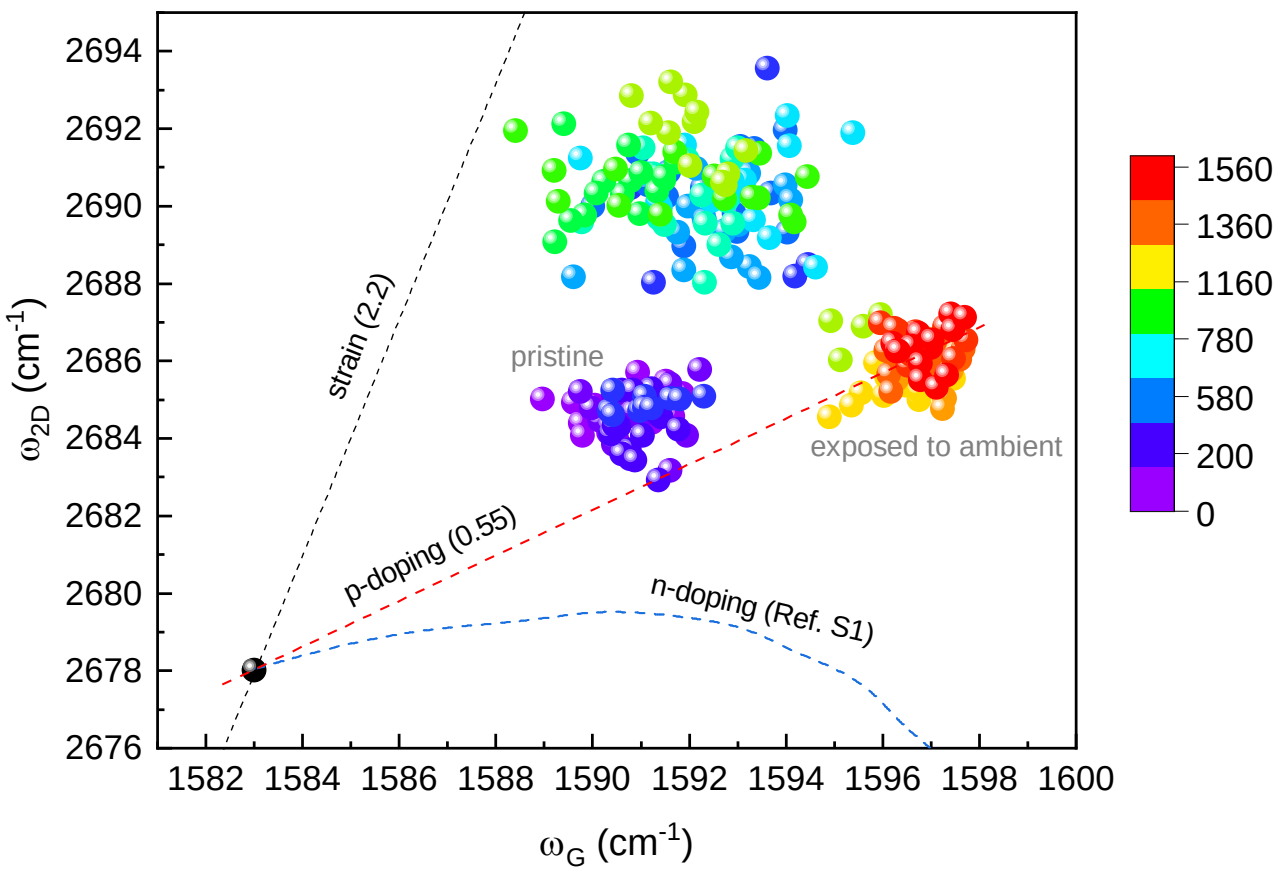

***Supplementary Figure S5: G- vs 2D-mode plot after correction for the effect of disorder according to experimental data from Supplementary Ref. S5.***

The data from the Supplementary Fig. S3 suggests that the pristine graphene in our experiments is initially pre-strained and p-doped. After the second plasma exposure, large disorder is introduced leading to the increased strain and slight n-doping of the sample. In contrast, after the exposure to ambient, the strain is released and the sample is strongly p-doped probably due to adsorption from ambient.

We have also confirmed the disorder effect by performing additional measurements on the sample B (Fig. S4), where we get identical results. We also note, that the defect density is affected by the charge carrier density[S4,8]. Therefore, in Fig. 2a in the main text we accounted for this effect and corrected the defect density values accordingly.

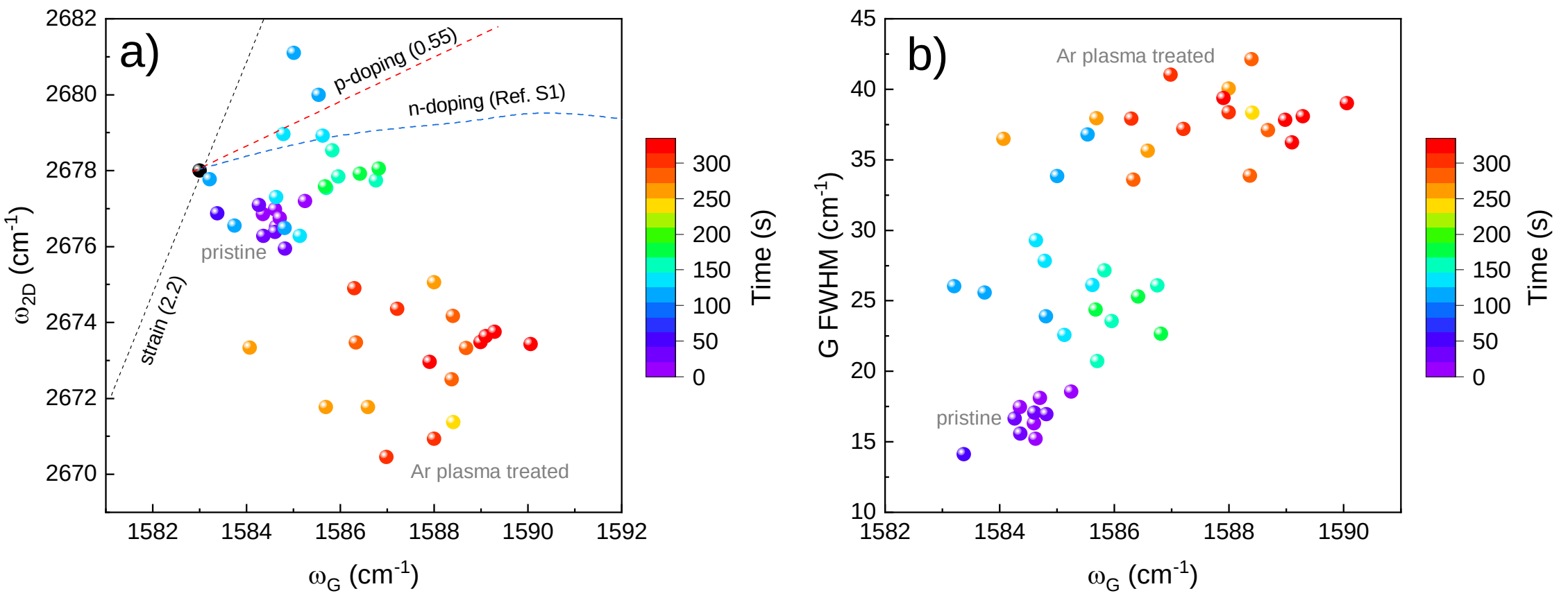

***Supplementary Figure S6: Time-resolved analysis of the Raman spectra for the Sample B. a)*** *$\omega_G$ vs. $\omega_{2D}$ plot and* ***b)*** *$\omega_G$ vs. G FWHM plot. Strain and doping induced at higher defect densities correspond to the behavior of Sample A. Defect density for the sample B is reaching $3\times10^{12}$ cm$^{-2}$.*

## SLOW DYNAMIC UPON AIR EXPOSURE

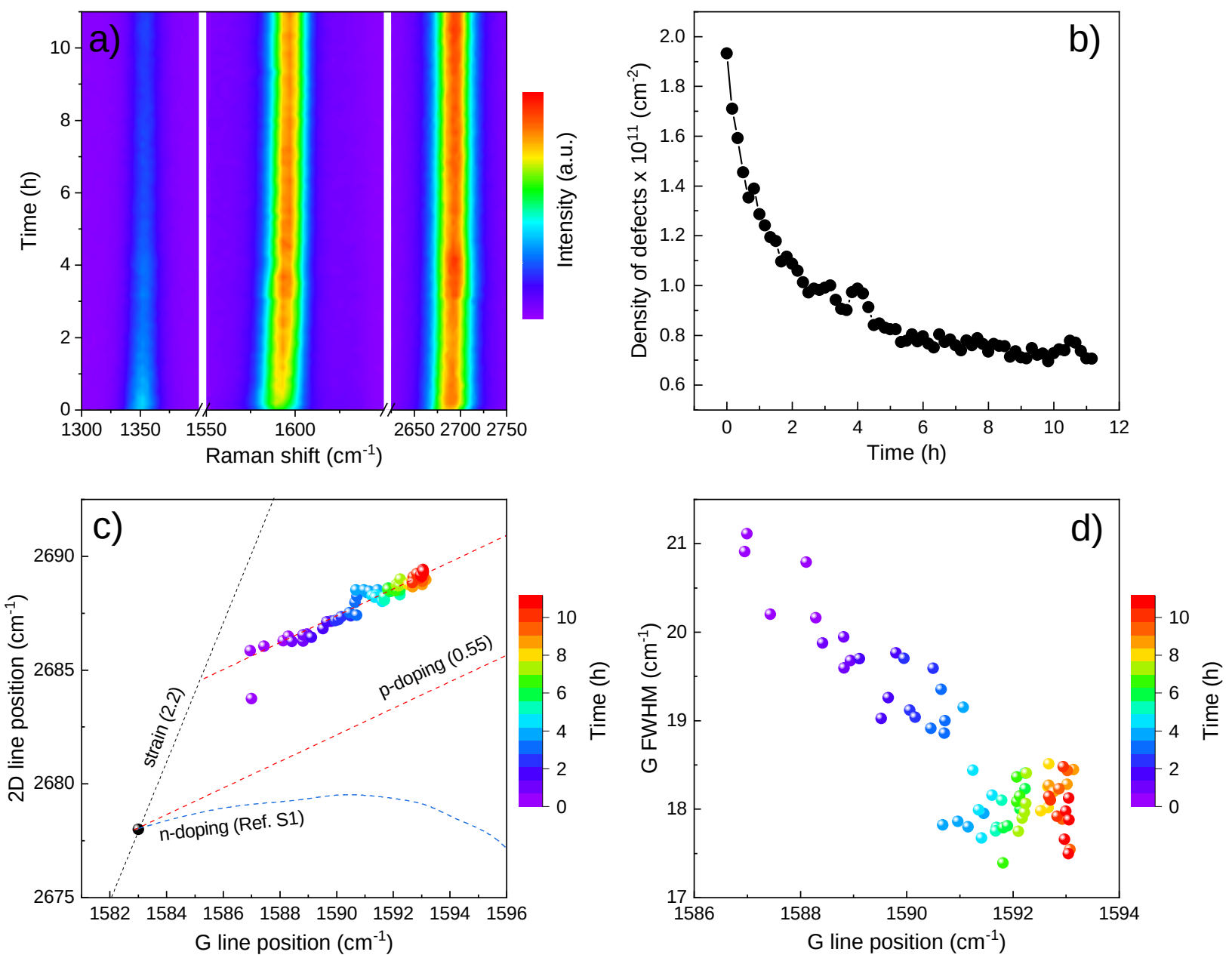

*Supplementary Figure S7: Slow dynamic over the 11 h during the exposure to air in Sample C: **a)** color map of Raman spectra, **b)** density of defects, **c)** $\omega_G$ vs. $\omega_{2D}$ plot and **d)** $\omega_G$ vs. G FWHM plot. The figure clearly shows the induction of p-doping and decrease of the defect density in the sample C over 11 h of the exposure to ambient.*

## DFT CALCULATIONS

Finite-size graphene nanoflakes are used to model graphene to avoid large supercells. This approach has been successfully adopted to model graphene and its nanostructures in previous works[S9–15]. The size of the graphene flakes ($C_{150}H_{30}$) ensures a reliable analysis of binding energy, charge carrier density, and strain, which are the key quantities monitored herein.

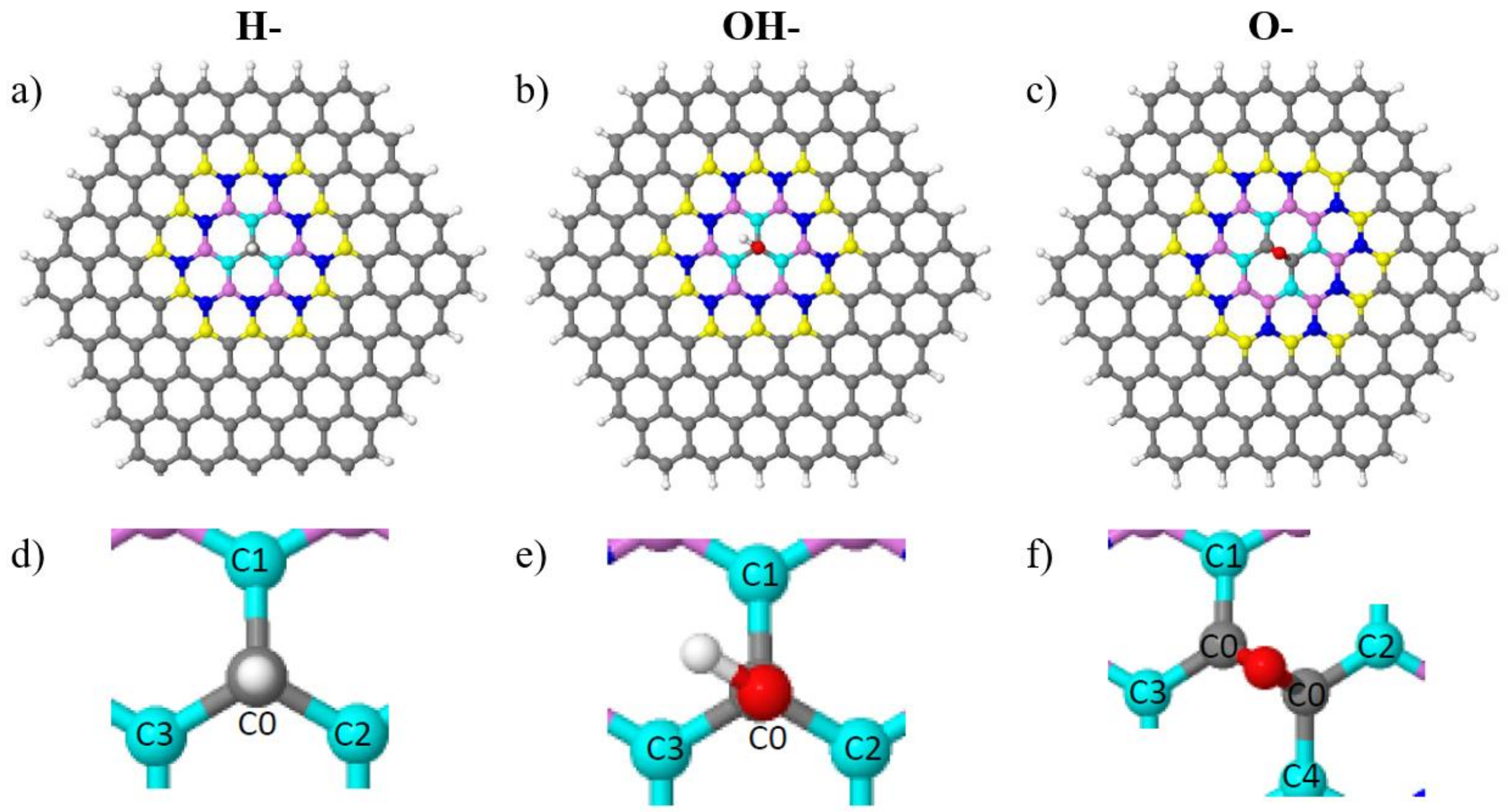

***Supplementary Figure S8: The pictures of H- (a, d); OH- (b,e) and O-defects (c,f) for which the DFT calculation of binding energies, charge carrier density and strain are summarized in the Table S1.***

**Supplementary Table S1:** Binding energies, strain and charge carrier density calculated using DFT for different types of functional defects in graphene.

| **Functional defect type** | **Binding energy, eV** | **Strain, %** | **Charge carrier density, × $10^{12}$ $cm^{-2}$** |
|---|---|---|---|
| H- | -0.839 | 0.2 | 7.1 (n-doping) |
| OH- | -0.993 | 0.2 | -6.5 (p-doping) |
| O- | -2.584 | 0.2 | -11.9 (p-doping) |

The data obtained in Supplementary Table S1 includes the analysis of strain and charge carrier densities up to the 4th nearest neighbors from the carbon atom with the corresponding functional defect as shown in Supplementary Fig. S6.

## SUPPLEMENTARY REFERENCES